# Electron Irradiation of Crystalline Nitrous Oxide Ice at Low Temperatures: Applications to Outer Solar System Planetary Science

Duncan V. Mifsud,[1,†] Sándor Góbi,[2,3] Péter Herczku,[1] Béla Sulik,[1] Zoltán Juhász,[1] Sergio Ioppolo,[4] Nigel J. Mason,[1,5] and György Tarczay[2,3,6,†]

1 HUN-REN Institute for Nuclear Research (Atomki), Debrecen H-4026, Hungary

2 MTA-ELTE Lendület Laboratory Astrochemistry Research Group, Institute of Chemistry, Eötvös Loránd University, Budapest H-1518, Hungary

3 Laboratory of Molecular Spectroscopy, Institute of Chemistry, Eötvös Loránd University, Budapest H-1518, Hungary

4 Centre for Interstellar Catalysis, Department of Physics and Astronomy, Aarhus University, Aarhus DK-8000, Denmark

5 Centre for Astrophysics and Planetary Science, School of Physics and Astronomy, University of Kent, Canterbury CT2 7NH, United Kingdom

6 Centre for Astrophysics and Space Science, Eötvös Loránd University, Budapest H-1518, Hungary

† Corresponding authors: Duncan V. Mifsud (mifsud.duncan@atomki.hu)
György Tarczay (tarczay@chem.elte.hu)

## ORCID Identification Numbers

| | | | |
|---|---|---|---|
| Duncan V. Mifsud | 0000-0002-0379-354X | Zoltán Juhász | 0000-0003-3612-0437 |
| Sándor Góbi | 0000-0002-7039-8099 | Sergio Ioppolo | 0000-0002-2271-1781 |
| Péter Herczku | 0000-0002-1046-1375 | Nigel J. Mason | 0000-0002-4468-8324 |
| Béla Sulik | 0000-0001-8088-5766 | György Tarczay | 0000-0002-2345-1774 |

## Abstract

The radiation chemistry and physics of solid $N_2O$ have been increasingly studied due to its potential presence on the surfaces of cold, outer Solar System bodies. However, to date, no study has investigated systematically the influence of temperature on this chemistry and physics. In this present study, crystalline $N_2O$ ices were irradiated using 2 keV electrons at five different temperatures in the 20-60 K range and the radiolytic dissociation of the molecular solid (as well as the radiolytic formation of seven product molecules) was quantified through the *G*-value. Our results indicate that temperature does indeed play a role in the radiolytic destruction of crystalline $N_2O$, with higher temperatures being associated with higher destruction *G*-values. The formation *G*-values of NO, $NO_2$, $N_2O_2$, $N_2O_3$, $N_2O_4$, $N_2O_5$, and $O_3$ were also noted to vary with temperature, with each product molecule exhibiting a distinct trend. The applications of our experimental results to further understanding solid-phase radiation chemistry in the outer Solar System are discussed.

## 1. Introduction

The confirmed detection of a number of simple inorganic nitrogen-bearing molecules (e.g., $N_2$ and $NH_3$) on the surfaces of several outer Solar System dwarf planets and moons such as Pluto, Triton, Umbriel, and Eris **[1-4]** has led to an increased interest in the solid-phase astrochemistry of $N_2O$. This is in part due to the fact that the surfaces of these bodies are routinely exposed to ionising radiation in the form of galactic cosmic rays and the solar wind, which is known to engender chemical changes in astrophysical ices. Furthermore, laboratory studies have repeatedly demonstrated that the irradiation of ices made up of nitrogen-containing molecules mixed with $H_2O$ or other oxygen-bearing molecules (e.g., $O_2$, CO, or $CO_2$) that may also co-exist on the surfaces of these icy bodies leads directly to the formation of $N_2O$ **[5-16]**. It is to be noted, however, that the presence of $N_2O$ itself has yet to be confirmed on the surfaces of outer Solar System moons and dwarf planets.

Nevertheless, the repeated formation of $N_2O$ in laboratory irradiations of ices representative of the surfaces of outer Solar System planetary-like bodies is suggestive of this molecule being a likely surface component. Accordingly, the radiation physics and chemistry of solid $N_2O$ under conditions relevant to astrochemistry have been considered by a number of previous studies that have shown that several molecules, including nitrogen oxides of the type $N_xO_y$ (where $x \leq y$) and $O_3$, are produced as a result of the radiolytic dissociation of $N_2O$ and subsequent radical recombination reactions **[17-25]**. However, many of these experimental astrochemistry studies have been performed under temperature conditions more relevant to the cold, dense interstellar medium (i.e., 10-20 K), and the radiolytic physics and chemistry of $N_2O$ in the outer Solar System (which is characterised by higher temperatures in the range of 30-60 K) only have been inferred from the results of these lower temperature studies.

For instance, the study of Almeida *et al.* **[19]** was able to quantify the destruction cross-section of solid $N_2O$ during its irradiation at 10 K by 1.5 MeV $N^+$ ions, as well as the formation cross-sections of a number of radiolytic products. Using this calculated destruction cross-section together with the known flux of medium-mass ions at the orbit of Pluto **[26]**, Almeida *et al.* **[19]** estimated the half-life of solid $N_2O$ in the outer Solar System to be $9\times10^6$ years. However, such an estimation is critically reliant on the destruction cross-section measured by Almeida *et al.* **[19]** at 10 K not varying significantly with increasing temperatures. Despite this requirement, previous astrochemical investigations into the possible temperature dependence of $N_2O$ ice radiolysis (and, by the same token, the possible temperature dependence of the formation of radiolytic product molecules) have been few and far between.

Indeed, to the best of our knowledge, only two recent studies focused on the temperature dependence of the radiolysis of $N_2O$. The first, conducted by de Barros *et al.* **[18]**, investigated the irradiation of solid $N_2O$ at 11 and 75 K using 90 MeV $Xe^{23+}$ ions and calculated higher destruction cross-sections for the higher temperature experiment. This was attributed to the higher mobility of radiolytically derived radicals at higher temperatures, as well as the relevant reaction activation energy barriers being more easily overcome. The second study was that of Fulvio *et al.* **[20]**, who investigated the irradiation of $N_2O$ ice at 16 and 50 K using 0.2 MeV protons. Interestingly, the results of this latter study did not detect any noticeable differences between the radiolytic destruction rate of $N_2O$ at these two temperatures.

Although these two studies have made interesting contributions to our understanding of the role of temperature in the radiation-induced dissociation of solid $N_2O$ under conditions relevant to astrochemistry, a number of questions remain unresolved. These primarily relate to the method by which the $N_2O$ ices considered by de Barros *et al.* **[18]** and Fulvio *et al.* **[20]** were prepared, including the resultant solid phase of the ice. In principle, $N_2O$ ices prepared by condensation of a gas onto a cooled substrate at temperatures below *circa* 30 K should result in an amorphous structure, while those prepared in a similar fashion at higher temperatures should produce a crystalline ice **[27]**. This would suggest that each of the studies of de Barros *et al.* **[18]** and Fulvio *et al.* **[19]** would have considered one amorphous ice and one crystalline ice. However, it is well known that the phase adopted by $N_2O$ ices prepared by condensation of the gas on a cooled substrate is heavily dependent on the rate of ice deposition, with higher deposition rates resulting in crystalline structures even at low temperatures (for a more complete discussion of this phenomenon, see the works of Hudson *et al.* **[27]** and Gerakines and Hudson **[28]**). Since neither of these two previous studies defined the rate at which their ices were grown, there is some ambiguity regarding the phases of the ices that were irradiated in those studies.

Knowledge of the phase of an irradiated astrophysical ice analogue is important, as there has been an increasing body of literature in recent times that has demonstrated that simple molecular ices (including $N_2O$) are prone to higher radiolytic decay rates if in an amorphous phase compared to a crystalline one **[25,29,30]**. As such, if the role of temperature on the radiolytic destruction of $N_2O$ is to be accurately quantified, then it is necessary to exclude any potential influence that the phase of the target ice may bear upon the process. The aim of the present study, therefore, was to systematically and quantitatively explore the influence of temperature on the radiation physics and chemistry of solid $N_2O$, in the absence of any contributing effects from other factors such as the phase adopted by the ice. This has been achieved by preparing five purely crystalline ices under identical conditions and then irradiating them using 2 keV electrons at five different temperatures (i.e., 20, 30, 40, 50, and 60 K) that are relevant to the outer Solar System. Further details on the experimental apparatus and techniques used in this study are provided in Section 2, while the experimental results and their implications for astrochemistry in the outer Solar System are discussed in Section 3. Finally, concluding remarks are given in Section 4.

## 2. Methods

Experiments were performed using the Ice Chamber for Astrophysics-Astrochemistry (ICA); a laboratory set-up for radiation astrochemistry research located at the HUN-REN Institute for Nuclear Research (Atomki) in Debrecen, Hungary. The set-up has been described in detail in other publications **[31,32]**, and so only a brief overview of the most salient features will be provided here. The ICA (Figure 1) is an ultrahigh-vacuum-compatible stainless-steel chamber maintained at a base pressure of a few $10^{-9}$ mbar by the combined action of a scroll pump and a turbomolecular pump. Within the centre of the chamber is a gold-coated oxygen-free high-conductivity copper sample holder into which a number of ZnSe deposition substrates may be installed. The sample holder and deposition substrates may be cooled to a minimum temperature of 20 K using a closed-cycle helium cryostat, although it is possible to accurately control the temperature over a range of 20-300 K.

Astrophysical $N_2O$ ice analogues were prepared *via* background condensation of the gas onto ZnSe substrates cooled to 60 K at an average rate of 0.04 μm min$^{-1}$. The selected temperature and ice growth rate were sufficiently high to ensure that the resultant $N_2O$ ices were purely crystalline structures **[27]**. $N_2O$ ice growth was monitored *in situ* using Fourier-transform mid-infrared transmission absorption spectroscopy over a wavenumber range of 4000-650 cm$^{-1}$ and at a spectral resolution of 1 cm$^{-1}$. Mid-infrared spectra of a prepared ice may be used to quantify the $N_2O$ column density $N$ (molecules cm$^{-2}$) by measuring the integrated absorbance $S$ (cm$^{-1}$) of a characteristic absorption band. These quantities are related as per Eq. 1:

$$N = 2.303\frac{S}{A} \tag{1}$$

where $A$ is the so-called integrated band strength constant (cm molecule$^{-1}$) of that particular absorption band. The column density of a prepared astrophysical ice analogue is related to its thickness $d$ (μm) as per Eq. 2:

$$d = 10{,}000\frac{NM}{\rho N_{\mathrm{A}}} \tag{2}$$

where $M$ is the molar mass of the ice (g mol$^{-1}$), $\rho$ is the ice density (g cm$^{-3}$), and $N_A$ is the Avogadro constant ($6.02\times10^{23}$ molecules mol$^{-1}$). In the case of crystalline $N_2O$ ice, $M$ = 44 g mol$^{-1}$ and $\rho$ = 1.591 g cm$^{-3}$ **[27]**.

Once prepared to a satisfactory thickness, the crystalline $N_2O$ ices were subsequently cooled to the desired temperature and individually exposed to a 2 keV electron beam having an average particle flux of $5\times10^{13}$ electrons cm$^{-2}$ s$^{-1}$, with projectile electrons impacting the ices at an angle of 36° to the surface normal (Figure 1). The electron beam flux, homogeneity, and beam spot size had been defined prior to commencing the irradiations using the method described previously by Mifsud *et al.* **[32]**. Each ice was irradiated for a total of 30 minutes, corresponding to an average delivered fluence of $9\times10^{16}$ electrons cm$^{-2}$ and an average administered dose of $1.38\times10^{9}$ Gy **[33]**. Several mid-infrared absorption spectra were acquired over the course of each irradiation, thus allowing for any possible temperature dependence of the radiation physics and chemistry of $N_2O$ ice to be quantitatively probed. A summary of the experiments performed is given in Table 1.

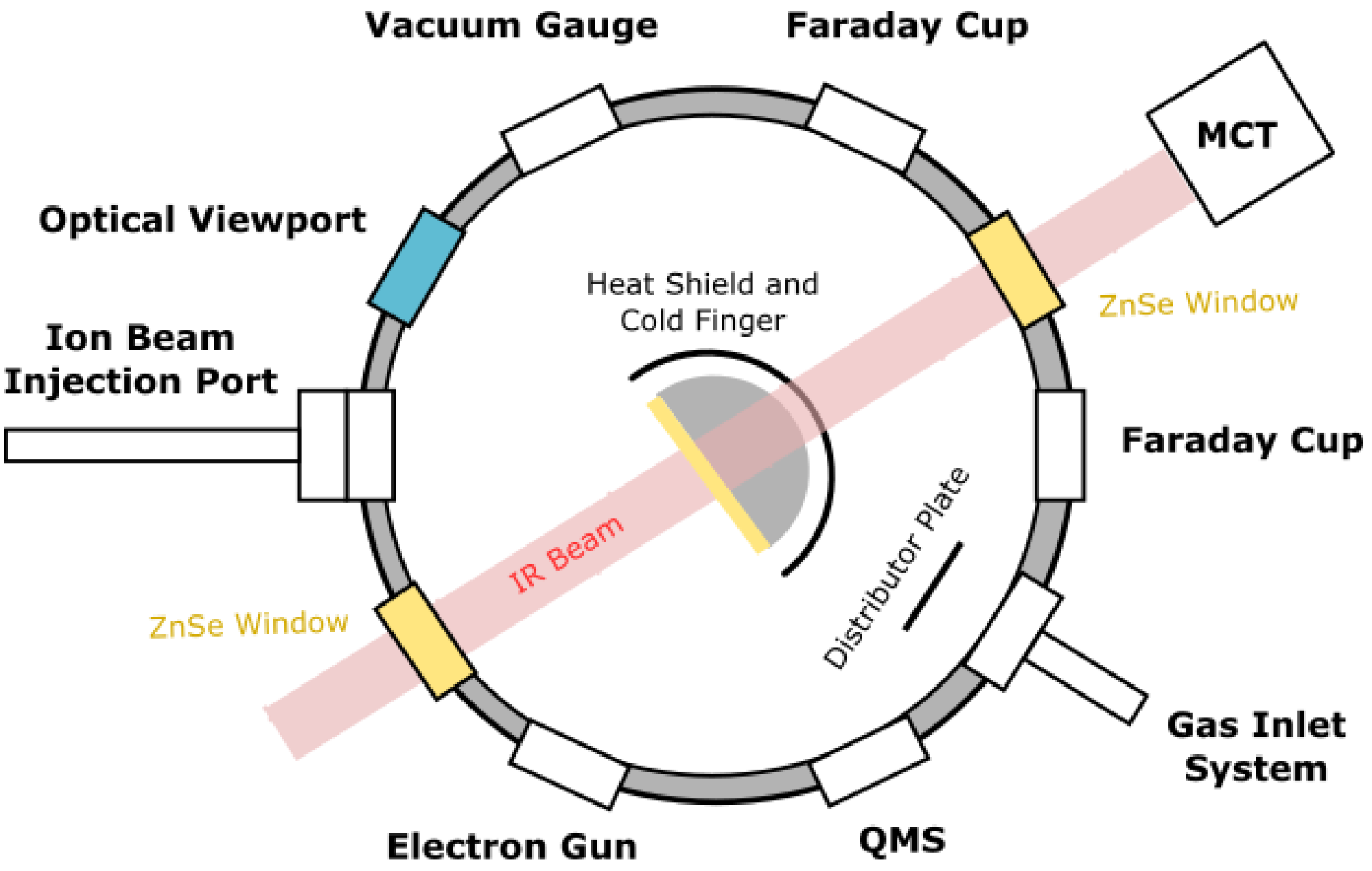

**Figure 1.** Top-view schematic diagram of the ICA. Irradiation experiments are nominally performed using the arrangement depicted here, where the sample holder is maintained orthogonal to the mid-infrared spectroscopic beam and projectile electrons impact the deposited ices at an angle of 36° to the normal. Image reproduced from the work of Mifsud *et al.* **[32]** with the kind permission of *The European Physical Journal*.

**Table 1.** Summary of the experiments performed in this study. Information on the temperature at which each ice was irradiated ($T_{irrad}$), its initial column density ($N_0$) and thickness ($d$), as well as the electron fluxes ($\Phi$), fluences ($\phi$), and doses ($\delta$) used in each irradiation experiment is provided.

| Ice | $T_{irrad}$ (K)[a] | $N_0$ ($10^{17}$ cm$^{-2}$)[b] | $d$ (μm) | $\Phi$ ($10^{13}$ cm$^{-2}$ s$^{-1}$) | $\phi$ ($10^{16}$ cm$^{-2}$) | $\delta$ ($10^{9}$ Gy) |
|---|---|---|---|---|---|---|
| 1 | 20 | 4.01 | 0.18 | 5.27 | 9.49 | 1.45 |
| 2 | 30 | 5.64 | 0.26 | 5.29 | 9.52 | 1.46 |
| 3 | 40 | 6.20 | 0.29 | 4.88 | 8.78 | 1.34 |
| 4 | 50 | 3.96 | 0.18 | 4.75 | 8.55 | 1.31 |
| 5 | 60 | 5.60 | 0.26 | 4.75 | 8.55 | 1.31 |

[a] Although the ices were irradiated at different temperatures, they were all prepared at 60 K.
[b] Calculated by measuring the $\nu_3$ band at 2237 cm$^{-1}$ over an integration range of 2264-2222 cm$^{-1}$ and taking $A$ to be $5.21\times10^{-17}$ cm molecule$^{-1}$ **[27]**.

The radiolytic destruction of the $N_2O$ ices considered in this study as a result of their interaction with a 2 keV electron beam has been quantified through the *G*-value, defined as the number of $N_2O$ molecules destroyed per 100 eV of energy deposited into the ice. This is calculated by plotting the measured column density of $N_2O$ (in molecules $cm^{-2}$) as a function of electron fluence in dose units (i.e., in units of eV $cm^{-2}$ rather than electrons $cm^{-2}$). The initial slope of the decreasing $N_2O$ column density multiplied by 100 yields the *G*-value. The radiolytic synthesis of a number of product molecules was also quantified using the *G*-value, which was determined in an analogous way considering the initial slope of the increasing product molecule column density. When considering the *G*-values of astrophysical ice analogues undergoing radiolytic dissociation or synthesis in this way, it is important to only consider the initial fluence points early on in the irradiation where radiolytic dissociation or synthesis is likely the only process contributing to changes in the measured column density of the molecule of interest. Measurements at these fluence points are likely to be far from the steady-state, where radiolytic destruction and synthesis are balanced through the contribution of 'back reactions' to give a near-constant abundance at high fluences.

It is also important to note that the calculation of *G*-values from mid-infrared absorption spectra assumes that the only mechanism contributing to the loss of $N_2O$ molecules in the irradiated astrophysical ice analogues is radiation-induced molecular dissociation. In fact, it is possible that intact $N_2O$ molecules may be ejected into the gas phase as a result of sputtering. Such an effect has been extensively documented when using large ions as projectiles **[34-36]**, and has also been documented when using electrons **[37,38]**, albeit to a more limited extent. To the best of our knowledge, the sputtering of $N_2O$ ice at low temperatures by incident electrons has not yet been studied experimentally. Since mid-infrared spectroscopic measurements of $N_2O$ ice column density variations during irradiation cannot discriminate between losses due to true radiolytic destruction and those due to sputtering, it is necessary to emphasise that the *G*-values calculated in this present study for $N_2O$ radiolysis therefore represent upper bounds to these parameters. That being said, it is likely that our calculated values for these parameters are close to the true values since previous experimental studies have determined that electron-induced sputtering, although non-negligible, is not a major contributor to material loss during irradiation **[34-43]**.

## 3. Results and Discussion

### 3.1 Mid-Infrared Absorption Spectroscopy and Radiation Chemistry

Mid-infrared absorption spectra of solid $N_2O$, both before and after irradiation using 2 keV electrons, are shown in Figure 2 and band assignments of the parent and product molecules are provided in Tables 2 and 3, respectively. All ices exhibited the anticipated absorption features associated with crystalline $N_2O$ after deposition at 60 K, which did not vary significantly in their wavenumber position or profile after having been cooled to the desired irradiation temperature (Table 2). Moreover, the band assignments of the parent $N_2O$ ices reported herein are in good agreement with those of previous studies **[44,45]**. Upon the onset of irradiation, the absorption bands attributed to $N_2O$ were observed to decline while new bands attributable to various product molecules emerged against the background continuum.

Among the radiolytic products were several nitrogen oxides of the type $N_xO_y$ (where $x \leq y$), including NO, $NO_2$, $NO_3$ (tentative detection), $N_2O_2$, $N_2O_3$, $N_2O_4$, and $N_2O_5$. Additionally, $O_3$ and $N_3$ were also detected (Figure 2). The formation of these products as a result of the irradiation of solid $N_2O$ has been reported by several previous studies **[16-25]**, and is presumed to proceed primarily as a result of the radiolytic fission of the O–N bond in the $N_2O$ molecule to yield $N_2$ and a free, supra-thermal oxygen atom (note that $N_2$ cannot be detected spectroscopically among the radiolytic products due to it being a homonuclear diatomic species, and thus infrared-inactive). The preference for the fission of this bond as opposed to the N≡N bond in the $N_2O$ molecule lies in the lower bond energy of the former (1.7 eV) compared to that of the latter (4.9 eV) **[46,47]**.

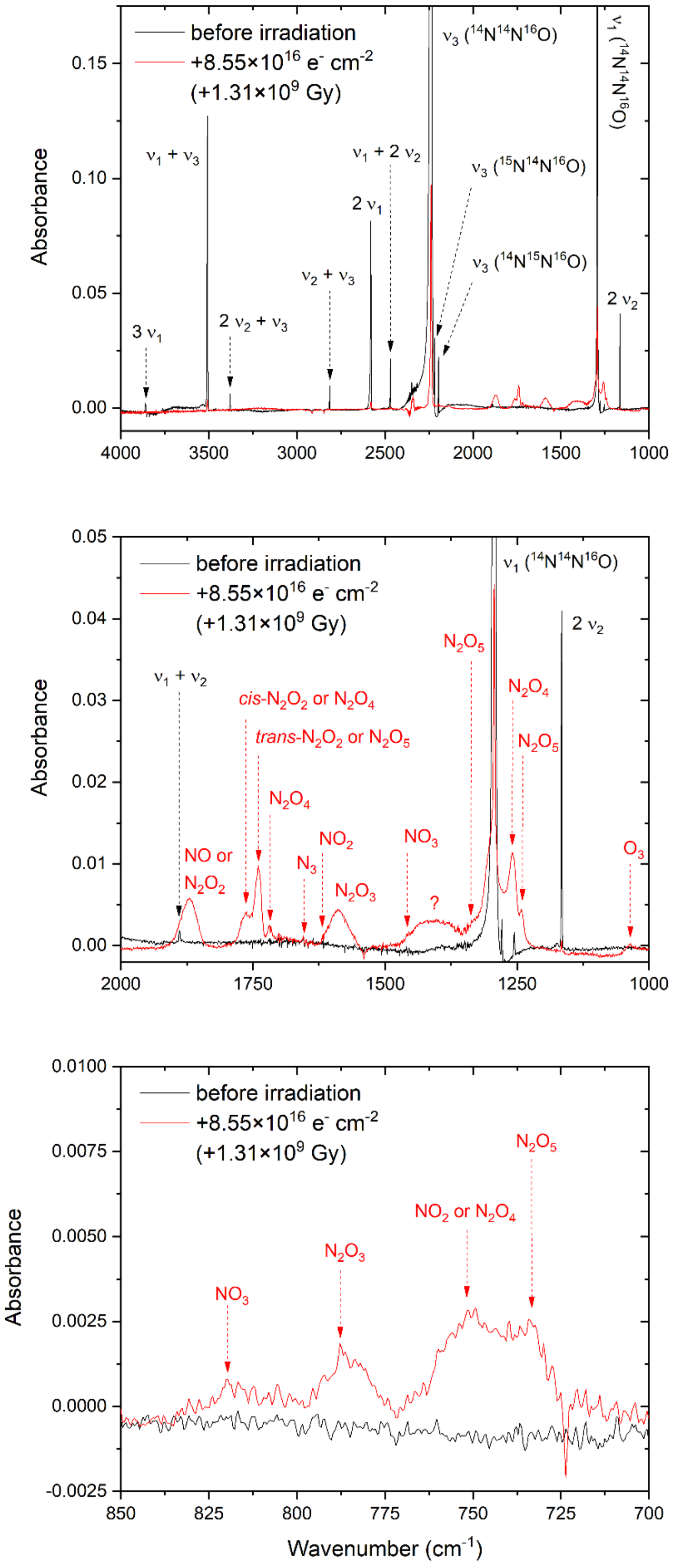

**Figure 2.** Mid-infrared absorption spectra of $N_2O$ ice before (black trace) and after (red trace) irradiation using 2 keV electrons at 60 K. Absorption bands of the parent ice are labelled in black, while absorption bands of the radiolytic products are labelled in red.

**Table 2.** Mid-infrared band assignments of pristine $N_2O$ ice at different experimental temperatures. Note that all ices considered in this study were crystalline ices prepared by the background condensation of the gas at 60 K followed by cooling to the desired temperature.

| $N_2O$ Band | Wavenumber Position ($cm^{-1}$) | | | | | | |
|---|---|---|---|---|---|---|---|
| | 20 K Ice | 30 K Ice | 40 K Ice | 50 K Ice | 60 K Ice | Reference Study[a] | Reference Study[b] |
| 2 $\nu_2$ | 1165.3 | 1165.3 | 1165.3 | 1165.3 | 1165.3 | 1165.6 | 1166.6 |
| $\nu_1$[†] | 1255.0 | 1255.5 | 1255.5 | 1255.1 | 1255.0 | 1256.5 | 1252.5 |
| $\nu_1$[‡] | 1278.6 | 1278.8 | 1278.8 | 1278.6 | 1278.6 | 1279.8 | 1276.1 |
| $\nu_1$[*] | 1293.1 | 1293.2 | 1293.2 | 1293.1 | 1293.1 | 1293.4 | 1291.2 |
| $\nu_1 + \nu_2$ | 1889.4 | 1889.4 | 1889.0 | 1889.5 | 1889.5 | 1891.0 | 1885.3 |
| $\nu_3$[⁑] | 2195.6 | 2195.6 | 2195.6 | 2195.6 | 2195.6 | 2195.8 | 2188.8 |
| $\nu_3$[‡] | 2219.3 | 2219.3 | 2219.3 | 2219.3 | 2219.3 | 2219.7 | 2212.9 |
| $\nu_3$[*] | 2237.6 | 2237.6 | 2237.6 | 2237.5 | 2237.6 | 2238.0 | 2235.6 |
| $\nu_1 + 2\,\nu_2$ | 2468.9 | 2469.3 | 2469.0 | 2469.1 | 2469.1 | 2473.4 | 2466.5 |
| 2 $\nu_1$ | 2580.5 | 2580.8 | 2580.8 | 2580.3 | 2580.4 | 2579.5 | 2575.3 |
| $\nu_2 + \nu_3$ | 2814.6 | 2814.7 | 2814.7 | 2814.3 | 2814.2 | 2813.7 | 2809.3 |
| $2\,\nu_2 + \nu_3$ | 3380.2 | 3380.2 | 3380.2 | 3379.7 | 3380.0 | - | 3374.5 |
| $\nu_1 + \nu_3$ | 3508.9 | 3509.0 | 3509.1 | 3508.6 | 3508.4 | 3508.0 | 3499.3 |
| 3 $\nu_1$ | 3861.9 | 3861.9 | 3861.3 | 3861.9 | 3861.9 | - | 3853.7 |

[a] Taken from the study of Dows **[44]** which considered a crystalline $N_2O$ ice at 70 K.
[b] Taken from the study of Łapinski et al. **[45]** which considered a crystalline $N_2O$ ice at 6 K.
[†] This band is associated with the $^{14}N^{14}N^{18}O$ isotopologue.
[‡] This band is associated with the $^{15}N^{14}N^{16}O$ isotopologue.
[*] This band is associated with the $^{14}N^{14}N^{16}O$ isotopologue.
[⁑] This band is associated with the $^{14}N^{15}N^{16}O$ isotopologue.

**Table 3.** Mid-infrared band assignments of molecular products resulting from the 2 keV electron irradiation of $N_2O$ ice at different temperatures. Note that all $N_2O$ ices considered in this study were crystalline ices deposited at 60 K and subsequently cooled to the desired irradiation temperature.

| Product | | Position ($cm^{-1}$) | | | | | References |
|---|---|---|---|---|---|---|---|
| Molecule | Mode | 20 K Ice | 30 K Ice | 40 K Ice | 50 K Ice | 60 K Ice | |
| $N_2O_5$ | $\nu_{11}$ | 735.6 | 735.8 | 735.0 | 733.2 | 734.0 | Refs. **[61,62]** |
| $NO_2$ or $N_2O_4$ | $\nu_2$ or $\nu_{12}$ | 751.8 | 753.2 | 751.6 | 752.8 | 749.8 | Refs. **[61,63,64]** |
| $N_2O_3$ | $\nu_4$ | 785.9 | 784.9 | 784.7 | 784.9 | 787.9 | Ref. **[49]** |
| $NO_3$ | 2 $\nu_4$ | - | - | 820.1 | 818.3 | 820.1 | Ref. **[60]** |
| $O_3$ | $\nu_3$ | 1038.1 | 1039.9 | 1037.1 | 1037.5 | 1037.1 | Ref. **[65]** |
| $N_2O_5$ | $\nu_{10}$ | 1243.4 | 1242.8 | 1242.0 | 1241.6 | 1241.8 | Ref. **[6]** |
| $N_2O_4$ | $\nu_{11}$ | 1260.1 | 1260.5 | 1259.9 | 1258.9 | 1258.3 | Ref. **[6]** |
| $N_2O_5$ | $\nu_2$ | 1340.6 | 1339.0 | 1339.8 | 1339.2 | 1340.4 | Refs. **[62,66]** |
| ? | ? | 1413.2 | 1416.8 | 1421.2 | 1407.7 | 1406.5 | - |
| $NO_3$[†] | $\nu_3$ | 1457.0 | 1456.8 | 1456.9 | 1457.1 | 1456.8 | Refs. **[6,58,59]** |
| $N_2O_3$ | $\nu_3$ | 1595.1 | 1594.5 | 1591.5 | 1589.9 | 1589.7 | Ref. **[61]** |
| $NO_2$ | $\nu_3$ | 1612.7 | 1612.7 | 1613.3 | 1612.5 | 1612.3 | Refs. **[61,63]** |
| $N_3$ | $\nu_3$ | 1653.2 | 1652.8 | 1652.6 | 1653.4 | 1653.0 | Ref. **[6]** |
| $N_2O_4$ | $\nu_5$ | 1717.4 | 1717.2 | 1716.4 | 1717.5 | 1717.2 | Refs. **[61,67]** |
| *trans*-$N_2O_2$ or $N_2O_5$ | $\nu_5$ or $\nu_1$ | 1740.0 | 1739.8 | 1741.0 | 1740.0 | 1740.0 | Refs. **[6,17]** |
| *cis*-$N_2O_2$ or $N_2O_5$ | $\nu_5$ or $\nu_9$ | 1764.4 | 1764.2 | 1763.6 | 1764.2 | 1763.8 | Ref. **[6]** |
| NO or $N_2O_2$ | $\nu_1$ | 1868.8 | 1867.5 | 1868.6 | 1868.4 | 1868.7 | Refs. **[61,68-71]** |

[†] The assignment of this band is tentative. See text for more details.

Once liberated, free supra-thermal oxygen atoms may combine with one another to sequentially yield $O_2$ (which is not detectable in acquired mid-infrared spectra due to it being infrared-inactive) and $O_3$ **[19,48-50]**. Liberated oxygen atoms may also react with intact $N_2O$ molecules, resulting in either the formation of $N_2O_2$ **[17,51]** or, alternatively, in the dissociation of $N_2O$ to $N_2$ and $O_2$ **[50,51]**. $N_2O_2$ itself may undergo unimolecular decomposition to yield two NO molecules **[17]**. The formation of NO represents the starting point for very rich nitrogen chemistry in the irradiated ice due to the large number of reactions in which this molecule may participate (Figure 3). Aside from its dimerisation to form $N_2O_2$, NO may also react with an oxygen atom to produce $NO_2$ **[49,50]**. Presumably, the formation of $NO_3$ results from the addition of atomic oxygen to $NO_2$, although it is likely that this reaction occurs in competition with that yielding NO and $O_2$ **[19,49,50,54]**.

An alternative formation pathway towards NO is the radiation-induced dissociation of $N_2O$ to NO and atomic nitrogen, although this dissociation pathway is less likely to occur than that yielding $N_2$ and atomic oxygen as discussed earlier. Nevertheless, the release of supra-thermal nitrogen atoms throughout the ice allows for them to participate in atom addition reactions and thereby contribute to the diversity of nitrogen-bearing products in the irradiated ice. Perhaps the most evident example of such a reaction is the sequential synthesis of $N_2$ and $N_3$ **[55]**, in an analogous fashion to the synthesis of $O_3$ as a result of oxygen atom addition reactions. Furthermore, the addition of atomic nitrogen to $NO_2$ may result in its dissociation to two NO molecules or, alternatively, in the formation of $N_2O$ and atomic oxygen **[19,49,50,54]**; while its addition to NO results in the formation of $N_2$ and atomic oxygen **[49,50]**.

Regarding the formation of the nitrogen oxides containing two nitrogen atoms that were detected in our irradiated ices, a step-wise formation of oxygen atom addition – starting from $N_2O$ and leading to the sequential formation of $N_2O_2$, $N_2O_3$, $N_2O_4$, and $N_2O_5$ – is certainly possible. However, previous studies have suggested that other reactions might be more efficient in producing these molecules. This may be related to the fact that each step-wise addition of atomic oxygen occurs in direct competition with the dissociation of the nitrogen oxide to $N_2$ and $O_2$ **[19]**. Instead, the formation of $N_2O_3$ was proposed to result from the reaction between NO and $NO_2$, or from the reaction between $N_2O$ and $O_2$ **[19,49,50]**. The dimerisation of $NO_2$ has been suggested to be the principal route towards the formation of $N_2O_4$, although the reaction between $N_2O$ and $O_3$ may also contribute towards its formation **[19,49,50]**. Lastly, $N_2O_5$ is thought to result from one of several reaction processes involving three species, such as that involving $N_2O$ and two $O_2$ molecules, that involving $O_3$ and two $NO_2$ molecules, and that involving $N_2O$, $O_3$, and atomic oxygen **[17,19]**. The complexity of the reactions leading to the formation of $N_2O_5$ explains why previous studies have reported it to be among the lowest abundance products formed after the irradiation of $N_2O$ ice **[17-20]**.

Although practically all the mid-infrared absorption bands depicted in the spectra shown in Figure 2 could be assigned and the chemistry leading to the formation of the molecules with which they are associated elucidated by this and previous studies, we wish to draw attention to one feature in particular that we have not been able to unambiguously assign. Upon inspection of the middle panel in Figure 2, it is possible to note a broad absorption band ranging over approximately 1460-1360 $cm^{-1}$ which we have labelled using a '?'. This feature is present in all our spectra of irradiated $N_2O$ ice, irrespective of the temperature at which the irradiation was performed. There currently exists little consensus in the literature as to the origin of this band and, indeed, not all previous studies have observed it after the irradiation of solid $N_2O$.

The studies of de Barros *et al.* **[18]** and Pereira *et al.* **[21]** both observed this band in their recorded mid-infrared spectra, but did not explicitly assign it to any particular molecule or vibrational mode. In their study, Fulvio *et al.* **[20]** observed this band over a wavenumber range of about 1500-1350 $cm^{-1}$ (peaking at approximately 1430 $cm^{-1}$) and assigned it to $NO_3$. However, this assignment is challenging to confirm. Previous gas-phase **[56-58]** and matrix-isolation **[59]** spectroscopic studies have consistently placed the $\nu_3$ mode of $NO_3$ at higher wavenumbers of about 1492 $cm^{-1}$, and this has been largely corroborated by computational work **[60]**. Moreover, to the best of our knowledge, no previously published spectroscopic study concerned with $NO_3$ has described the $\nu_3$ mode as a broad band as suggested by the assignment of Fulvio *et al.* **[20]**. It is thus difficult to assign the broad band observed in our spectra over the 1460-1360 $cm^{-1}$ wavenumber range to the $\nu_3$ mode of $NO_3$ with any certainty. However, a small shoulder is apparent towards the higher wavenumber end of this broad band at about 1457 $cm^{-1}$. Given the better (although still somewhat unsatisfactory) match between the position of this band and the literature value of the $NO_3$ $\nu_3$ mode, we have tentatively assigned this shoulder to $NO_3$ while leaving the broad band unassigned. Further work clarifying the exact nature of the broad band at approximately 1460-1360 $cm^{-1}$ is required.

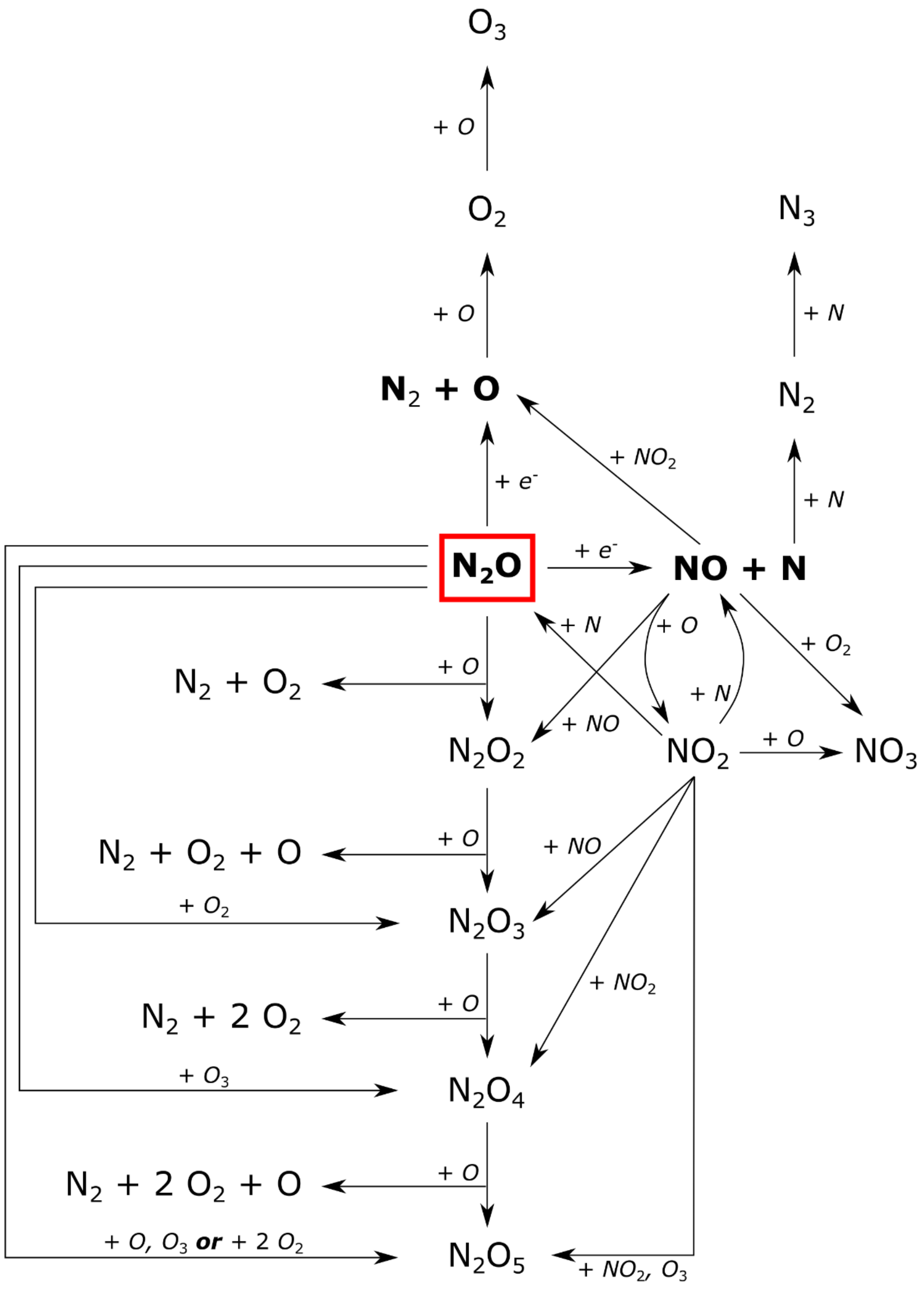

**Figure 3.** Reaction network detailing the chemical transformations occurring within electron irradiated $N_2O$ ices. The primary electron-induced dissociation pathways of the $N_2O$ parent molecule are indicated in bold.

### 3.2 $N_2O$ Radiolytic Destruction as a Function of Temperature

One immediately striking result of our quantitative analysis is the evidence of more extensive radiolytic destruction of crystalline $N_2O$ at higher temperatures. Indeed, by comparing plots of the normalised column density of this molecule as a function of electron fluence or radiation dose, it is possible to note that the most extensive decay occurs at the highest investigated temperatures of 50 and 60 K, while the lowest degree of destruction is observed in ices irradiated at 20 and 30 K (Figure 4). The measured $N_2O$ column density of the ice irradiated at 40 K appears to decay at an intermediate rate compared to those irradiated at lower and higher temperatures (Figure 4). The *G*-values associated with the radiolytic destruction of crystalline $N_2O$ by 2 keV electrons were also computed for each investigated temperature (Figure 5, Table 4) and, as expected from the data plotted in Figure 4, the *G*-values became more negative with increasing temperature thus indicating greater radiation-induced molecular destruction at these higher temperatures. The relationship between temperature and *G*-value could be modelled reasonably well by a linear fit given by the equation:

$$G_{(2\ \mathrm{keV\ e^-})}(\text{crys. N}_2\text{O}) = -0.05T_{\text{irrad}} - 0.5 \tag{7}$$

Our results are therefore in broad agreement with the previous work of de Barros *et al.* **[18]**, who irradiated $N_2O$ ice at 11 and 75 K using 90 MeV $Xe^{23+}$ ions and found that the destruction cross-section for the higher temperature irradiation was approximately one and a half times greater than that for the lower temperature irradiation. However, for a number of reasons, the results of that study do not allow for a direct comparison of the temperature dependence of the radiolytic destruction of $N_2O$ ice. Firstly, there is uncertainty as to the nature of the phases of solid $N_2O$ investigated by de Barros *et al.* **[18]**. Indeed, upon examination of their pre-irradiation spectra (their Figures 2-4), it is possible to notice certain similarities in the mid-infrared spectra of the ices acquired at 11 and 75 K; most notably the fact that the absorption features in the lower temperature ice's spectrum are sharper than those expected for a purely amorphous ice as demonstrated by Hudson *et al.* **[27]**. This would imply that the lower temperature ice studied by de Barros *et al.* **[18]** was most likely some admixture of amorphous and crystalline solid which probably arose due to the use of a fast deposition rate. Given the known influence of phase on the radiolytic decay rate of an astrophysical ice analogue **[25,29,30]**, it is not possible to conclusively state that the higher radiolytic destruction rate of $N_2O$ observed by de Barros *et al.* **[18]** at 75 K is solely due to the influence of temperature. Secondly, it is to be noted that the higher temperature irradiation experiment of de Barros *et al.* **[18]** was performed fairly close to the sublimation temperature of $N_2O$ **[18,72]**. As such, it is possible that some of the losses of $N_2O$ may have arisen from sublimation rather that true radiolysis.

Nevertheless, our systematic study conducted on pure crystalline ices at temperatures well below the sublimation point has confirmed that the irradiation of $N_2O$ at higher temperatures does indeed result in a higher radiolytic destruction rate, as was concluded by de Barros *et al.* **[18]**. This is likely due to the increased mobility through the ice matrix of the radicals derived from the radiolysis of $N_2O$, which thus decreases the likelihood of 'back reactions' leading to its reformation. Our results, however, contrast with those of Fulvio *et al.* **[20]**, who considered the irradiation of $N_2O$ ice at 16 and 50 K using 0.2 MeV protons but did not record any noticeable differences in the rates of radiolytic decay at these two temperatures. The reason for the discrepancy between our results and those of Fulvio *et al.* **[20]** is not immediately obvious, and thus it is recommended that future studies seeking to further elucidate the temperature dependence of the radiolytic destruction of $N_2O$ ice make use of different charged particles across a wide energy range.

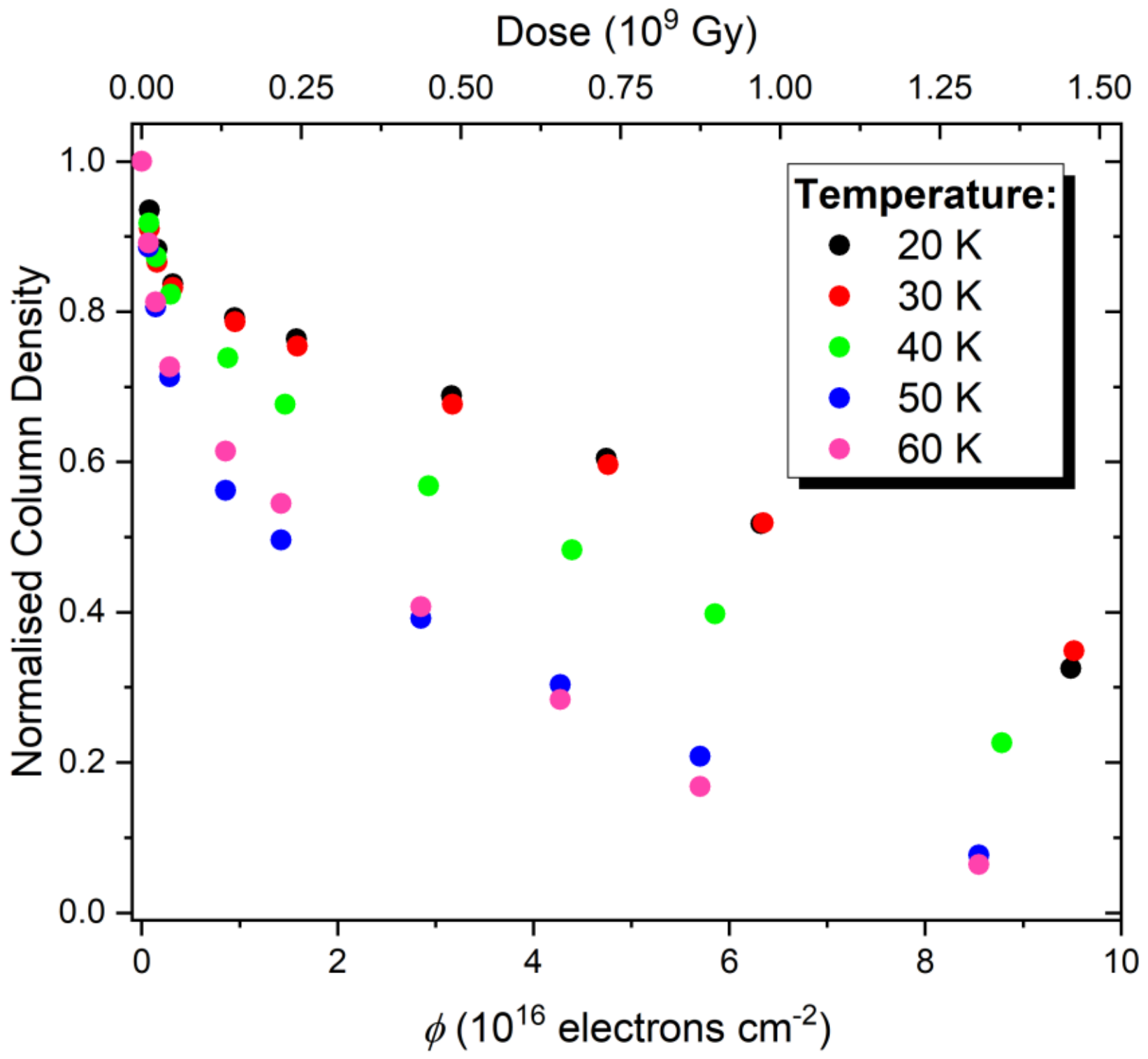

**Figure 4.** Normalised column density of crystalline $N_2O$ measured as a function of 2 keV electron fluence. Error bars have been omitted for the sake of clarity, but uncertainties in the normalised column density measurements are on the order of 7%.

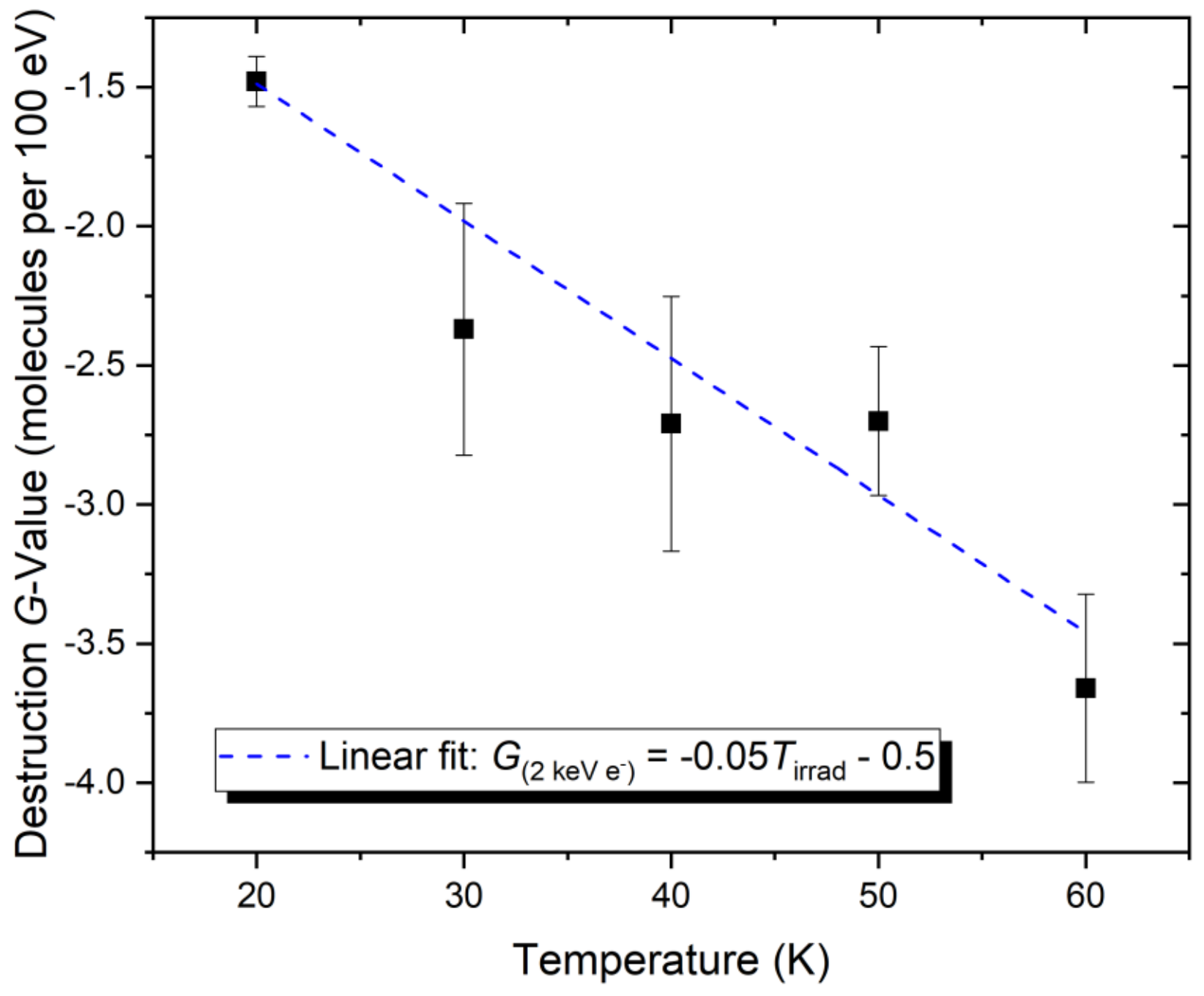

**Figure 5.** Temperature dependence of the $G$-value for solid crystalline $N_2O$ as a function of temperature during irradiation by 2 keV electrons.

**Table 4.** Calculated *G*-values for the 2 keV electron irradiation of $N_2O$ at different temperatures. Product molecule column densities were calculated using the integrated band strength constants reported in Jamieson *et al.* **[6]** and Stirling *et al.* **[73]** for the following bands: 1869 $cm^{-1}$ (NO), 1613 $cm^{-1}$ ($NO_2$), 1740 $cm^{-1}$ ($N_2O_2$), 785 $cm^{-1}$ ($N_2O_3$), 1260 $cm^{-1}$ ($N_2O_4$), 1243 $cm^{-1}$ ($N_2O_5$), and 1038 $cm^{-1}$ ($O_3$). Note that the 1869 $cm^{-1}$ band used to calculate the NO *G*-value may have some contribution from $N_2O_2$, although this is likely small at the early electron fluences that were measured. Note also that the absorption bands of $N_3$, though detectable in acquired mid-infrared spectra, were too small to allow for an accurate analysis of the *G*-value.

| **Molecule** | ***G*-Values** | | | | |
|---|---|---|---|---|---|
| | **20 K** | **30 K** | **40 K** | **50 K** | **60 K** |
| $N_2O$ | −1.48 ± 0.09 | −2.37 ± 0.45 | −2.71 ± 0.46 | −2.70 ± 0.27 | −3.66 ± 0.34 |
| NO | 0.901 ± 0.238 | 0.904 ± 0.109 | 0.957 ± 0.107 | 1.018 ± 0.064 | 1.292 ± 0.023 |
| $NO_2$ | 0.087 ± 0.013 | 0.083 ± 0.019 | 0.040 ± 0.017 | 0.023 ± 0.009 | 0.019 ± 0.006 |
| $N_2O_2$ | 0.168 ± 0.043 | 0.324 ± 0.063 | 0.285 ± 0.016 | 0.590 ± 0.078 | 0.753 ± 0.093 |
| $N_2O_3$ | 0.068 ± 0.084 | 0.150 ± 0.018 | 0.156 ± 0.041 | 0.199 ± 0.020 | 0.174 ± 0.024 |
| $N_2O_4$ | 0.014 ± 0.002 | 0.021 ± 0.004 | 0.027 ± 0.004 | 0.039 ± 0.005 | 0.061 ± 0.007 |
| $N_2O_5$ | 0.008 ± 0.001 | 0.011 ± 0.001 | 0.015 ± 0.001 | 0.019 ± 0.001 | 0.017 ± 0.001 |
| $O_3$ | 0.166 ± 0.030 | 0.279 ± 0.044 | 0.278 ± 0.059 | 0.250 ± 0.056 | 0.203 ± 0.061 |

### 3.3 Product Molecule Radiolytic Synthesis as a Function of Temperature

The irradiation of solid $N_2O$ using 2 keV electrons resulted in the synthesis of nine product molecules which could be detected in acquired mid-infrared spectra (Figure 2, Table 3): NO, $NO_2$, $NO_3$, $N_2O_2$, $N_2O_3$, $N_2O_4$, $N_2O_5$, $N_3$, and $O_3$. It is logical to assume that $N_2$ and $O_2$ were also formed as a result of the irradiation of $N_2O$, as shown in the reaction scheme in Figure 3. To complement our analysis of the variation of the $N_2O$ destruction *G*-value as a function of temperature, we have sought to quantify the temperature dependence of the formation *G*-value of these products. We have excluded $N_2$ and $O_2$ from this analysis due to their being infrared inactive; as well as $NO_3$ due to the detection of this molecule in our experiments being tentative. Moreover, the absorption bands associated with $N_3$ were rather small in our acquired spectra, and thus measurement of their integrated absorbances is associated with large uncertainties. For this reason, we have also excluded $N_3$ from our analysis.

#### *3.3.1 Nitrogen-Containing Products*

The *G*-values for the six nitrogen-containing products under investigation (i.e., NO, $NO_2$, $N_2O_2$, $N_2O_3$, $N_2O_4$, and $N_2O_5$) as a function of temperature are plotted in Figure 6. The variation in these *G*-values reveals that the radiation chemistry of crystalline $N_2O$ induced by 2 keV electrons is significantly impacted by the temperature at which the irradiation takes place. Considering first NO, which should primarily result from the radiolytic dissociation of $N_2O$ (Figure 3), it is possible to note that higher temperatures correspond to greater yields of this product molecule, with significant increases being noted on raising the temperature beyond 50 K. This aligns with the greater (i.e., more negative) destruction *G*-values of $N_2O$ at higher temperatures. The reason for this is that, at these temperatures, radiolytically generated radicals become more mobile within the ice matrix. This increased mobility allows them to diffuse away from the site of their formation, reducing the likelihood of the 'back reaction' that would lead to the reformation of $N_2O$.

A consequence of the larger yields of NO at higher irradiation temperatures is the greater abundance of $N_2O_2$, which results from the dimerisation reaction occurring between two NO molecules (Figure 3). The greater mobility of radicals at these higher temperatures allows for NO molecules to roam the ice matrix and thus increases the probability that they will come into contact with one another and react to yield $N_2O_2$. It is worthwhile noting that the computed formation *G*-values for $N_2O_2$, although lower than those corresponding to NO, are nevertheless fairly high. Indeed, the ratio between *G* ($N_2O_2$) and *G* (NO) increases from about 0.19 to about 0.58 as the irradiation temperature is increased from 20 to 60 K (Table 4). The greater rate of increase of the formation *G*-value of $N_2O_2$ with increasing temperature compared to that of NO may indicate that a significant fraction of radiolytically synthesised NO is sequestered into $N_2O_2$ at higher temperatures.

The relationship between the formation *G*-value and irradiation temperature is perhaps most stable for the $N_2O_3$ product, for which the *G*-value does not vary significantly from 0.15 (although it should be noted that the calculated *G*-value for the irradiation at 20 K is accompanied by comparatively larger uncertainties). A number of reasons could contribute to this relative stability of the *G*-value, including the lower reaction rate constants at low temperatures, increased rates of product molecule radiolysis at higher temperatures, and even perhaps the inherent instability of $N_2O_3$ itself **[20,74,75]**.

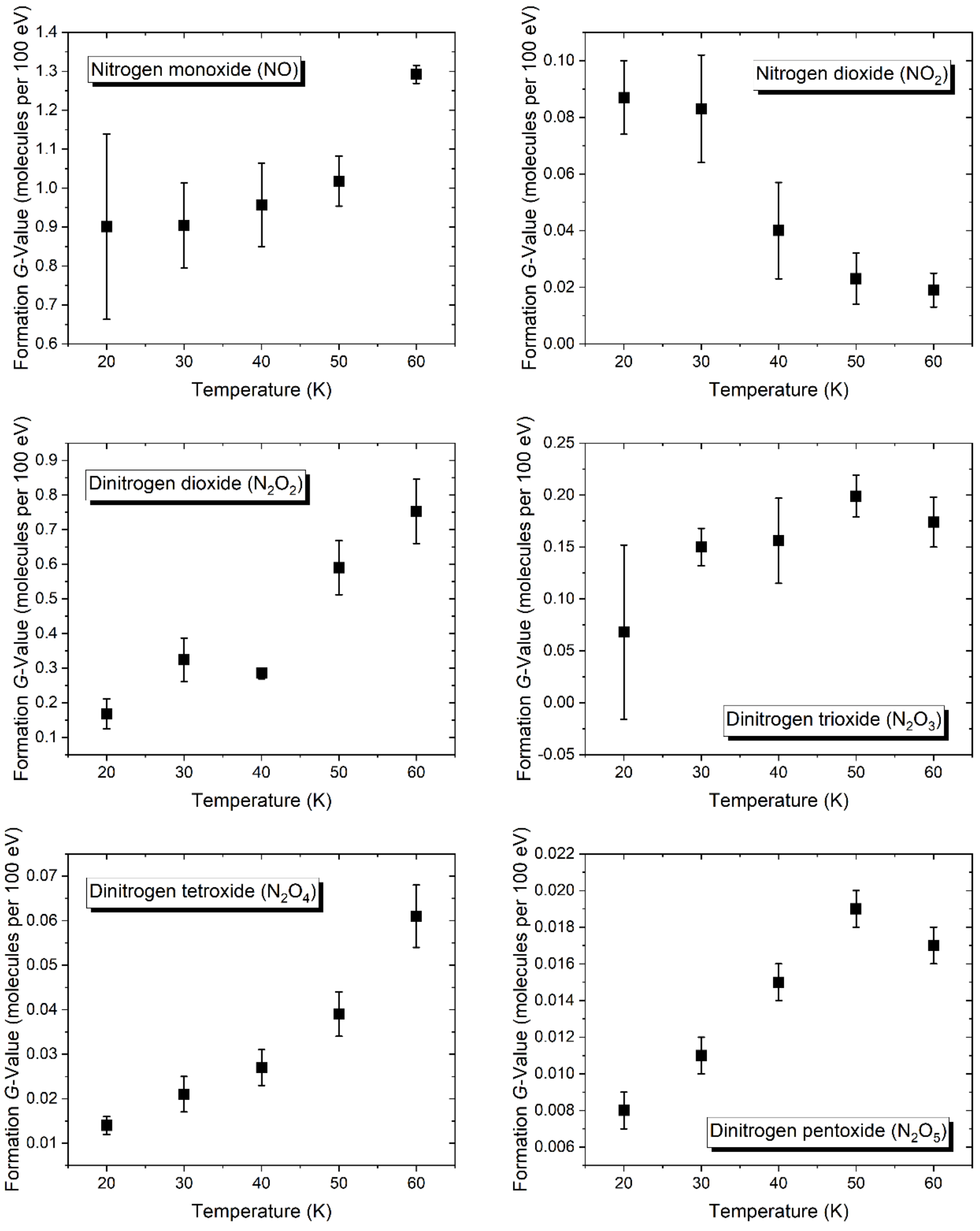

**Figure 6.** Temperature dependence of the *G*-value for nitrogen-containing products yielded from the 2 keV electron irradiation of crystalline $N_2O$ ice.

Interestingly, the relationship between the formation *G*-value of $NO_2$ and irradiation temperature is strikingly different, with the *G*-value sharply decreasing with increasing temperature (Figure 6) and the greatest decreases being observed at temperatures of 40 K and above. This trend is the opposite of that observed for $N_2O_4$, for which higher *G*-values are observed during the irradiation of $N_2O$ ices at higher temperatures (Figure 6). The most parsimonious explanation for this observation is that, at higher temperatures, the mobility of the nascent $NO_2$ molecules is sufficiently high that the rate at which they are able to encounter each other and dimerise to yield $N_2O_4$ (Figure 3) leads to an accumulation of the latter species at the expense of the former. Indeed, the ratio between *G* ($N_2O_4$) and *G* ($NO_2$) varies from 0.16 to 3.21 across the 20-60 K temperature range.

Finally, the last nitrogen-containing product for which *G*-values were computed was $N_2O_5$. As can be seen from Figure 6, formation *G*-values for this product were very low, never exceeding 0.02 at any of the investigated irradiation temperatures. The reason for this is most likely the complexity of the reactions required to synthesise this molecule in the solid phase, which typically require three precursor molecules (Figure 3). Yields of $N_2O_5$ increased as the irradiation temperature was increased from 20 to 50 K; however, a slight decrease in the *G*-value was registered on increasing the temperature further to 60 K. Although it is difficult to pinpoint an exact reason for this observation, it is possible that this was due to the fact that some of the precursors required for its formation, such as $O_2$ and $O_3$, undergo increased rates of sublimation from the ice at this higher temperature **[65,72,76]**. Such a result would also explain the measured relationship between the formation *G*-value of $O_3$ and the irradiation temperature (see next sub-section for more details).

### *3.3.2 Ozone*

The evolution of the formation *G*-value of $O_3$ as a function of the irradiation temperature of solid $N_2O$ is shown in Figure 7. It is clear that, on increasing the temperature from 20 to 30 K, the *G*-value also increases; almost doubling in magnitude. This may be attributed to the greater mobility of oxygen atoms within the ice that were liberated as a result of the radiolytic dissociation of $N_2O$ molecules. These oxygen atoms come together and react to yield sequentially $O_2$ and $O_3$ **[77-79]** (Figure 3). However, on further increasing the irradiation temperature to 40 K and beyond, the $O_3$ formation *G*-value was observed to steadily decrease. This is perhaps somewhat counterintuitive, since higher temperatures should, in principle, allow for the oxygen atoms to be even more mobile throughout the ice matrix, thus allowing them to come into contact and yield $O_3$ more efficiently. However, $O_2$ (i.e., the direct precursor to $O_3$) is a particularly volatile species under ultrahigh-vacuum conditions and previous studies have demonstrated that pure $O_2$ sublimates efficiently at temperatures lower than 40 K, although this may be shifted to higher temperatures if $O_2$ is entrapped within less volatile ice components **[72,80]**. As such, it is likely that the radiolytic formation of $O_3$ in our higher temperature experiments occurred in direct competition with the sublimation of $O_2$ from the ice, thus limiting the amount of $O_3$ that could actually be formed. Similar reasons were invoked in recent studies of the implantation of reactive sulphur ions into $CO_2$ ice, which observed the formation of oxygen-rich products (e.g., $O_3$ and $SO_2$) at 20 K but not at 70 K **[81,82]**.

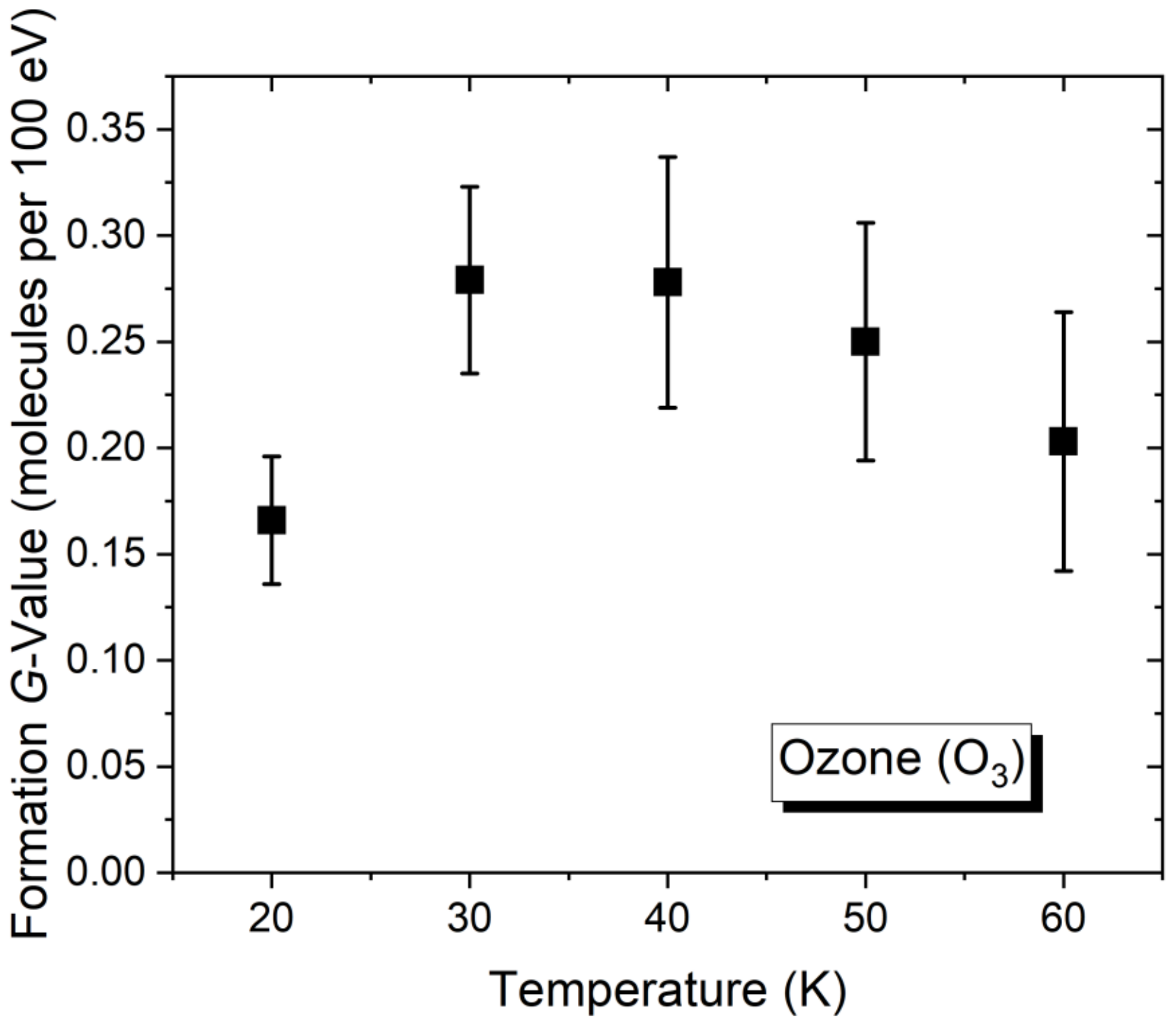

**Figure 7.** Temperature dependence of the *G*-value for $O_3$ yielded from the 2 keV electron irradiation of crystalline $N_2O$ ice.

### 3.4 Implications for Chemistry in the Outer Solar System

The icy surfaces of many bodies in the outer Solar System are known to be rich in nitrogen-containing species. Pluto's surface, for instance, is known to be dominated by $N_2$, with trace quantities of CO also being present **[1,83]**. The surface of Orcus is known to host $H_2O$ and $NH_3$ ices **[84,85]**, which are also thought to be present on the surface of Sedna **[86]**. Simple inorganic nitrogen-bearing ices have also been detected on the surfaces of other outer Solar System bodies, such as Triton, Umbriel, and Eris **[2-4]**. Ices containing both simple inorganic and more complex organic nitrogen-bearing species are also present on the surfaces of outer Solar System bodies in closer orbits of the Sun, such as a number of the icy satellites of Saturn **[87-90]**. Furthermore, it is now known that ammonium-rich salts are an important constituent of the icy nuclei of comets **[91-93]**.

Importantly, the surfaces of these outer Solar System bodies are all continuously exposed to ionising radiation in the form of the solar wind, galactic cosmic rays, and possibly, giant planetary magnetospheric plasma. Thus, although solid $N_2O$ has yet to be detected as a constituent of the icy surfaces of these small bodies, it is likely that it is indeed generated there as a result of the rich radiation chemistry that occurs between nitrogen- and oxygen-bearing molecules, as has been suggested by previous laboratory studies **[5-16]**. It is further suggested that, given the propensity of solid $N_2O$ to crystallise upon warming to temperatures greater than 30 K or upon its rapid condensation from the gas phase **[27]**, it is probable that should any substantial quantities of $N_2O$ ice exist on outer Solar System bodies then it would largely be crystalline; although it should be noted that ice amorphisation may occur to at least some extent in regions of high radiation fluxes **[94,95]**.

The surface temperatures of outer Solar System bodies vary significantly based on their orbital distances to the Sun, and it is also true that the exposure of icy material to fluxes of energetic ions, electrons, and ultraviolet photons can be greatly attenuated by shallow burial beneath other surface material. Our systematic computation of the $N_2O$ destruction *G*-values at relevant temperatures therefore allows us to effectively constrain the survivability of $N_2O$ on different icy Solar System worlds. Following the example of Zhang *et al.* **[96]**, we shall make the assumption that the survivability of a given solid-phase molecule in the outer Solar System can be approximated by considering the galactic cosmic ray ionising radiation dose rates in the Kuiper Belt at a distance of 40 AU from the Sun. The 2 keV electrons used in this study are actually very good mimics of galactic cosmic rays impacting the surfaces of outer Solar System bodies, since they have similar linear energy transfers to MeV ions within cryogenic molecular solids. However, it is to be noted that the chemistry triggered in ices irradiated by energetic (i.e., keV or MeV) ions and electrons is actually mediated by low-energy (i.e., <20 eV) secondary electrons released along the track of the primary irradiating particle, as documented by a number of previous studies **[97-99]**.

We have therefore made use of cosmic ray dose rates of $5.6\times10^{-3}$ and $1.6\times10^{-8}$ eV molecule$^{-1}$ year$^{-1}$ for icy material at the surface of such a Kuiper Belt Object (i.e., within the top $10^{-6}$ cm of the surface material) and for ices buried beneath $10^{-3}$ cm of surface material **[100,101]**, respectively. Therefore, by dividing our calculated $N_2O$ destruction *G*-values (Table 4) by 100 and inverting the resultant value, we can express the destruction rate of $N_2O$ at a given temperature in units of eV molecule$^{-1}$. Then, by dividing this resultant value by the quoted Kuiper Belt radiation dose rates, we can quantify the survivability of $N_2O$ as the equivalent number of years of exposure to cosmic radiation required to destroy it. This data is depicted in Figure 8.

A general trend is apparent in Figure 8 in which the survivability of crystalline $N_2O$ on an icy Kuiper Belt Object (as expressed through the number of years of exposure to cosmic radiation required to destroy it) decreases with increasing temperature. The greatest decrease in survivability appears to occur when increasing the surface temperature from 20 to 30 K, after which the survivability does not seem to change significantly upon further warming to 50 K. However, on warming to 60 K, another noticeable decrease in $N_2O$ survivability is registered. It is also evident that when $N_2O$ is present on the surface of the Kuiper Belt Object (as shown in panel (a) of Figure 8), its survivability is on the order of just a few thousand years. Conversely, when buried beneath just $10^{-3}$ cm of surface material, this survivability increases to a few billion years. Hence, it is possible to conclude that crystalline $N_2O$ ice may be preserved in the outer Solar System through astronomical timescales if it is shielded from incident ionising radiation by thin layers of surface material.

Although the results of our study indicate that crystalline $N_2O$ may be preserved on outer Solar System bodies if buried beneath $10^{-3}$ cm of surface material, it is important to note that our study is a somewhat oversimplified representation of processes actually occurring in the outer Solar System. In the first instance, it is likely that the abundance of $N_2O$ on any icy body in the outer Solar System is not only influenced by its destruction *via* its interaction with ionising radiation, but rather through a complex chemical network involving photochemical and photophysical processes, low-temperature thermal chemistry, radical combination reactions, sublimation and condensation cycles during surface warming events, and ejection to the gas phase during impacts; none of which has been considered in the present study.

Perhaps more importantly, however, is that solid $N_2O$ in surface and near-surface environments in the outer Solar System is likely a minor constituent in an ice mixture dominated by either $N_2$ or $H_2O$. The effect of this mixing on the destruction *G*-value of $N_2O$ is uncertain, since previous studies on the irradiation of nitrogen-bearing species mixed with other molecules have yielded varied results. For instance, Zhang *et al.* **[96]** demonstrated that the radiolytic destruction of pure $HOCH_2CH_2NH_2$ (i.e., ethanolamine) ice by 1 keV electrons at 20 K results in a more rapid decay compared to the irradiation of a 50:1 mixture of $H_2O$:$HOCH_2CH_2NH_2$. Conversely, Bordalo *et al.* **[102]** showed that $NH_3$ is more rapidly destroyed by irradiation with 536 MeV $Ni^{24+}$ ions if it is mixed with $H_2O$ than if it is irradiated as a pure ice. As such, it is recommended that future studies quantitatively assess the radiolysis of $N_2O$ in different ice mixtures relevant to the outer Solar System and compare this to the radiolysis of pure $N_2O$ ice. Nevertheless, our present study provides valuable insights into the fundamental radiation chemistry and physics of crystalline $N_2O$ at low temperatures, which will undoubtedly be of use when designing and interpreting the data from more complex experiments performed with the aim of elucidating the radiation chemistry of this molecule in the outer Solar System.

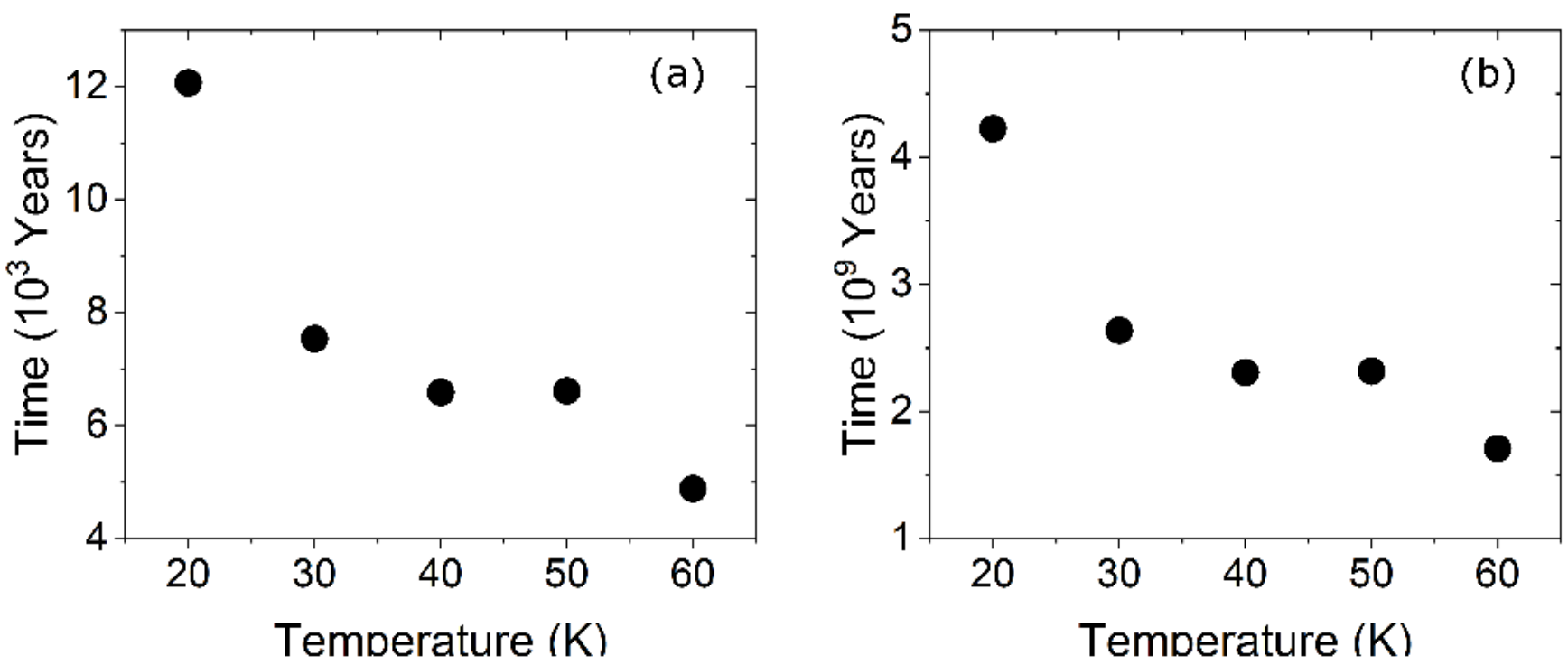

**Figure 8.** The temperature dependence of the theoretical number of years required to destroy $N_2O$ (a) at the surface (i.e., within the top $10^{-6}$ cm of surface material) of an icy Kuiper Belt Object and (b) buried beneath $10^{-3}$ cm of surface material.

## 4. Conclusions

In this study, we have systematically irradiated crystalline $N_2O$ ice using 2 keV electrons at five temperatures in the 20-60 K range with the aim of determining the effect of temperature on the radiolytic destruction of $N_2O$ and the formation of different radiolytic products. Our results have conclusively demonstrated that temperature does play a role in this chemistry, with higher temperatures being associated with higher $N_2O$ destruction *G*-values. Indeed, the relationship between the irradiation temperature and the destruction *G*-value can be modelled reasonably well by a linear function. The variation in the formation *G*-value of seven product molecules (i.e., NO, $NO_2$, $N_2O_2$, $N_2O_3$, $N_2O_4$, $N_2O_5$, and $O_3$) with temperature was also assessed. For most product molecules, increased temperatures resulted in increased formation *G*-values due to the greater mobility of radiolytically derived radicals through the ice matrix; although some exceptions to this trend were noted. Most notably, the formation *G*-value of $NO_2$ was found to substantially decrease with increasing temperature, likely as a consequence of its consumption in the synthesis of $N_2O_4$. More moderate decreases in the formation *G*-value with increasing temperature were recorded for $O_3$, probably due to the sublimation of its precursor molecule $O_2$ at higher temperatures. Finally, we have discussed the applicability of our results to $N_2O$ radiation chemistry on outer Solar System icy bodies, and have provided some tentative evidence for its preservation if buried beneath $10^{-3}$ cm of surface material, although more experimental work is required to validate this.

### Acknowledgements

The authors acknowledge support from the Europlanet 2024 RI which has been funded by the European Union's Horizon 2020 Research Innovation Programme under grant agreement no. 871149. The main components of the ICA set-up at Atomki were purchased using funds obtained from the Royal Society obtained through grants UF130409, RGF/EA/180306, and URF/R/191018. Further developments of the installation are supported in part by the Eötvös Loránd Research Network through grants ELKH IF-2/2019 and ELKH IF-5/2020. Support has also been received from the Research, Development, and Innovation Fund of Hungary through grant K128621. This paper is also based on work from the COST Action CA20129 MultIChem, supported by COST (European

Cooperation in Science in Technology). We are also thankful to Gergő Lakatos (University of Debrecen, Hungary) for his helpful comments and suggestions.

Sándor Góbi and Zoltán Juhász is grateful for the support of the Hungarian Academy of Sciences through the János Bolyai Research Scholarship. Sergio Ioppolo acknowledges support from the Danish National Research Foundation through the Centre of Excellence 'InterCat' (grant agreement no. DNRF150).

## Author Contributions

**Duncan V. Mifsud:** Conceptualisation, Methodology, Validation, Formal Analysis, Investigation, Data Curation, Writing – Original Draft, Writing – Review and Editing, Visualisation, Project Administration. **Sándor Góbi:** Methodology, Investigation, Writing – Review and Editing, Supervision. **Péter Herczku:** Validation, Investigation, Writing – Review and Editing, Supervision. **Béla Sulik:** Resources, Writing – Review and Editing, Project Administration, Funding Acquisition. **Zoltán Juhász:** Resources, Writing – Review and Editing, Project Administration, Funding Acquisition. **Sergio Ioppolo:** Conceptualisation, Resources, Writing – Review and Editing, Visualisation, Funding Acquisition. **Nigel J. Mason:** Conceptualisation, Resources, Writing – Review and Editing, Visualisation, Project Administration, Funding Acquisition. **György Tarczay:** Conceptualisation, Resources, Writing – Review and Editing, Visualisation, Project Administration, Funding Acquisition.

## Declaration of Interests Statement

The authors hereby declare that this work was performed in the absence of any interests (financial or otherwise) that could have biased the results or their interpretation.

## Graphical Abstract

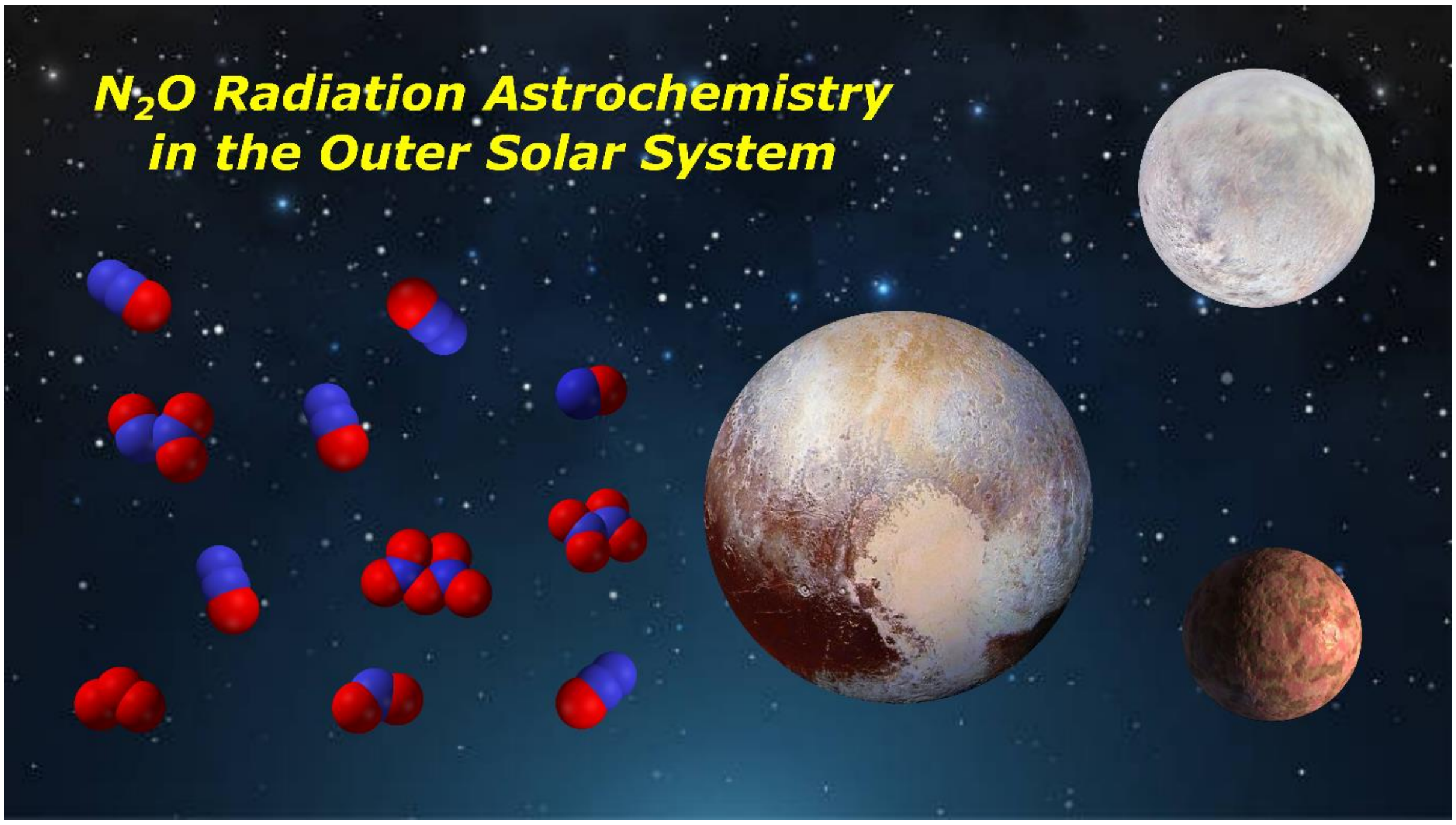